# Unveiling prethermalization and thermal processes through the simplest one-dimensional topological model

GUOWEN YANG[1,7], JIALE WANG[1,7], YICHUAN CHEN[1], LIMIN SONG[1], SHIQI XIA[1], DAOHONG SONG[1,2], ZHIGANG CHEN[1,2,5], AND NIKOLAOS K. EFREMIDIS[1,3,4,6]

[1]*MOE Key Laboratory of Weak-Light Nonlinear Photonics, TEDA Applied Physics Institute and School of Physics, Nankai University, Tianjin, 300457, China*

[2]*Collaborative Innovation Center of Extreme Optics, Shanxi University, Taiyuan, Shanxi 030006, China*

[3]*Department of Mathematics and Applied Mathematics, University of Crete, Heraklion, Crete 70013, Greece*

[4]*Institute of Applied and Computational Mathematics, FORTH, Heraklion, Crete 70013, Greece*

[5]*zgchen@nankai.edu.cn*

[6]*nefrem@uoc.gr*

[7]*These authors contributed equally to this work*

**Abstract:** Drawing on classical thermodynamic principles—such as the equipartition of energy and entropy maximization—extensive research has shown that the evolution of optical power in multimode optical systems tends toward a Rayleigh-Jeans distribution at thermal equilibrium. Understanding of the processes associated with the thermalization dynamics are of fundamental importance in analyzing and controlling such complex systems. In this work, we utilize a one-dimensional Su-Schrieffer-Heeger lattice as the simplest topological model to investigate the thermalization process of multiband systems in both topologically trivial and nontrivial regimes. Specifically, we identify that thermalization develops in three stages: (i) out-of-equilibrium dynamics, (ii) prethermal stage and (iii) final thermalization. Each individual band constitutes a subsystem that prethermalizes to the Rayleigh-Jeans distribution predicted from its power and internal energy. We find that this leads to a continuously varying prethermalization that eventually relaxes to the final thermal state (a dynamically evolving prethermal state). The presence of topological edge states can accelerate the thermalization process, although prethermal states exist both in the topologically trivial and nontrivial regimes. Factors such as bandgap width, temperature and nonlinearity that can influence the thermalization dynamics are examined in detail. Our work may offer valuable physical insights into understanding and controlling the thermalization process in multiband optical systems, paving the way for more efficient manipulation of light in complex settings.

## 1. Introduction

During the last few years, a significant amount of effort has been devoted in explaining the effect of beam self-cleaning[1-7]. Unexpectedly, as the energy of the pulse increases, the output profile transforms from a speckled pattern to a bell-shaped beam. Initially, different types of nonlinear multi-wave mixing processes have been suggested as the triggering mechanisms for this transition[8]. It was recently found that the origin of the beam self-cleaning effect lies in purely statistical principles that have been incorporated in the theory of optical thermodynamics[9], which has since triggered a surge of research interest[10-17]. With the formulation of statistical mechanics beyond the original multi-wave mixing paradigm as an explanatory framework, the theory of optical thermodynamics is directly linked with the classical thermodynamical principles. This includes the equipartition of energy, and the maximization of entropy in a microcanonical description. The presence of a weak nonlinearity merely enables interaction among eigenmodes - a necessary approach to achieve thermal equilibrium. Within the framework of optical thermodynamics, novel effects have been unveiled through extensive research efforts. This includes for example different types of entropic processes[18, 19], the optical pressure that is free from the complicated stress tensor calculations[20-22], and the resistance to thermalization caused by disorder or defects in lattice systems[23, 24]. Other examples include modulated and weighted thermalization distribution in non-Hermitian systems[25], and systems that conserve the optical orbital angular momentum[26, 27]. Experimentally, optical thermalization phenomena have been observed in multimode graded-index fibers[5, 27-29] and time synthetic mess lattices consisting of two coupled fiber loops[30]. In addition, very recently Joule thermal expansions have also been observed in nonlinear waveguide lattices by initially exciting a single waveguide in the photonic lattices[31]. Another interesting demonstration is the calorimetry measurements in graded index fibers demonstrate where the optical temperature and chemical potential dictate the flow of the energy and the power[32].

Understanding and taming the phenomena related to the thermalization processes is particularly important for applications. In this respect, thermalization effects have been studied in discrete optical settings[13, 23, 33, 34]. Although light can thermalize independently from the type of the nonlinearity involved, parameters such as the optical temperature and the chemical potential, the presence of bandgaps, the magnitude of the nonlinearity, and disorder can significantly affect the thermalization process. During the course of reaching thermal equilibrium, topological Haldane systems can get prethermalized in the trivial regime[35]. However, in the topologically nontrivial case, due to the presence of topological edge flow, prethermal states disappear. In addition, lattices with disorder or defects can also support prethermal states, where only a portion of the modes can get thermalized[24, 36].

On the other hand, the simplest one-dimensional topological model is the Su-Schrieffer-Heeger (SSH) lattice featuring a dimeric superlattice chain structure which was originally derived to describe the hopping of spinless fermion with alternating coupling coefficients[37]. Depending on the difference between the couplings, a band gap emerges, splitting the eigen spectra into two distinct bands. This gap is a crucial feature of topological systems, as it separates the bulk states from the edge or boundary states. The size of the gap often plays an important role in determining the robustness and stability of the topological states, as well as their sensitivity to perturbations. Locating at the edges of different topological domains, the SSH lattice can support two almost degenerate surface states between the two bands. In optics, such arrangement of waveguides have been extensively studied, including the observation and application of surface states[38-41], effects in the non-Hermitian regime[42-44] and nonlinearity-induced topological effects[45, 46]. However, the thermalization and prethermalization processes in such two-band systems remain unexplored, especially with respect to the impact of topological edges states, bandgap size, and nonlinearity.

In this work, we investigate the thermalization and prethermalization stages and processes involved in optical systems with bandgaps. Our model is the one-dimensional SSH lattice with cubic (Kerr) nonlinearity which, depending on the relative strength of the coupling coefficients, can be topologically trivial or nontrivial. Specifically, we find that the process starting from the initial out-of-equilibrium excitation to the final thermal state can be divided in three stages. The first one is the transition from the initial excitation condition to prethermalization. In the second (prethermal) stage, each band in isolation is characterized by dynamically evolving optical temperatures $T_j(z)$ and chemical potentials $\mu_j(z)$, where $j = 1,2$, and thus z-dependent Rayleigh-Jeans (RJ) distributions. In other words, we predict the existence of a dynamically evolving prethermal stage. This is in contrast to previous works, where prethermalization is connected to a single stationary distribution. In the third stage, at a certain thermalization distance, the values of the temperatures and the chemical potentials of the two bands converge to the same thermal values. We observe that band prethermalization emerges consistently in both topologically trivial and nontrivial regimes. However, in the topologically nontrivial regime, the presence of mid-gap surface states can reduce the thermalization distances, i.e., speed up the thermalization process. Thus, the mid-gap states act as thermal bridges between the two weakly coupled energy bands. We numerically compute the effect of the width of the bandgap, the strength of the Kerr nonlinearity, and the temperature on the thermalization and prethermalization distances. Interestingly, the prethermalization distances are always comparable for the two bands, and remain almost constant independently of the width of the bandgap. We expect our work can provide physical insights on understanding and taming the thermalization process of multiband optical settings.

## 2. Theory

The dynamics of an optical beam in waveguide array can be described by the paraxial light propagation equation, which has the same mathematical form as the Schrödinger equation with the addition of a cubic (Kerr) nonlinearity,

$$i\frac{\partial \phi}{\partial z} + \frac{1}{2k}\left(\frac{\partial^2 \phi}{\partial x^2} + \frac{\partial^2 \phi}{\partial y^2}\right) + V(x,y)\phi + \Gamma|\phi|^2\phi = 0. \tag{1}$$

In Eq. (1), $\phi$ describes the field of the optical wave, $x$ and $y$ are the transversal coordinates, $z$ is the propagation distance, $V(x,y)$ is the periodic refractive index distribution of the waveguides array, $\Gamma$ represents the Kerr nonlinear coefficient, and $k$ is the wavenumber. By applying the tight binding approximation, neglecting the higher-order coupling terms and counting only the nearest neighbor (NN) couplings, from Eq. (1), the following discrete nonlinear Schrödinger equation can be derived,

$$i\frac{d\phi_n}{dZ} + \sum_{\langle m,n \rangle} t_{mn}\phi_m + \gamma|\phi_n|^2\phi_n = 0, \tag{2}$$

where $\phi_n$ is the complex field amplitude on the $n$th site and $Z$ is the dimensionless propagating distance. In addition, $\langle m,n \rangle$ indicates NN lattice sites with coupling coefficient $t_{m,n}$, and $\gamma$ is the effective strength of the Kerr type nonlinearity. The nodal space lattice wavefunction $\phi_n$ can be expanded into the linear eigenmodes (or super-modes) $\phi_n = \sum_i c_i(z)\psi_n^{(i)}$ of Eq. (2), where $\varepsilon_i$ is the propagation constant of the $i$th eigenmode $\psi_n^{(i)}$, and $c_i(z)$ is the complex z-dependent amplitude. The optical power of mode $i$, or power occupation number, is defined by $n_i = |c_i|^2$. Equation (2) has two conservation laws, namely the Hamiltonian $H$ and the total power, which in modal space is given by $P = \sum_{i=1}^{M} n_i$. In the weakly nonlinear regime, the nonlinear part of the Hamiltonian, can be considered negligible and, thus, the internal energy can be approximated by its linear part $U = \sum_{i=1}^{M} \varepsilon_i n_i$, where $M$ is the number of eigenmodes. Note that discrete soliton formation indicates the transition from the weakly nonlinear to the strongly nonlinear regime. Furthermore, the optical entropy can be computed from the power occupation numbers through

$$S = \sum_{i=1}^{M} \ln n_i + M. \tag{3}$$

Following a microcanonical approach[9], we maximize the entropy subject to the two conserved quantities of the system $P$ and $U$, with $\alpha$ and $\beta$ being the Lagrange multipliers. Through calculations the power occupation numbers $\langle n_i \rangle$ are found to satisfy the following RJ distribution

$$\langle n_i \rangle = n_i^{RJ} = \frac{1}{\alpha + \beta \varepsilon_i} = \frac{T}{\varepsilon_i - \mu}, \tag{4}$$

where $T = 1/\beta$ and $\mu = -\alpha/\beta$ are the optical temperature and chemical potential, respectively. From Eq. (4), we see that for positive (negative) $T$ the lower (higher) order modes and mainly occupied. When $T = \pm\infty$ all the modes have the same power occupation numbers. On the other hand, the chemical potential $\mu = \partial U / \partial N$ it characterizes how the energy changes when power is added to the system. Utilizing the two conservation laws and the RJ distribution, it is easy to derive the following equation of state that relates the two conservation laws with the Lagrange multipliers of the system

$$U - \mu P = MT. \tag{5}$$

In this work, we investigate the one-dimensional SSH lattice. It is a diatomic lattice with alternating intracell and intercell coupling coefficients $t_1$ and $t_2$, respectively. For a finite size 1D SSH lattice with $M = 2N$ modes, under zero boundary conditions, the linear part of the Hamiltonian takes the form

$$H_L = t_1 \sum_{n=1}^{N} a_n^* b_n + t_2 \sum_{n=2}^{N} a_n^* b_{n-1} + \text{h.c.}, \tag{6}$$

where $a_n$ and $b_n$ indicate the optical field amplitudes located on sublattice a and b of the $n$th unit cell, and $N$ is the number of unit cells. Its bulk momentum space Hamiltonian can be written as

$$H(k) = \begin{pmatrix} 0 & t_1 + t_2 e^{-ik} \\ t_1 + t_2 e^{ik} & 0 \end{pmatrix} = d_1(k)\sigma_1 + d_2(k)\sigma_2, \tag{7}$$

where $\sigma_1$ and $\sigma_2$ are Pauli matrices, $k$ is the transverse wave vector, and $d_1(k) = t_1 + t_2 \cos k$, $d_2(k) = \sin k$. The winding numbers of the two bands are then given by $\nu_- = -\nu_+ = -\left(\frac{1}{\pi}\right) \int_0^{2\pi} [d\Phi(k)/dk] dk$ where $\Phi(k) = \arctan d_2(k)/d_1(k)$. Introducing the detuning between the intra- and inter-cell coupling $\delta = t_1 - t_2$, positive and negative signs of $\delta$ indicate the topological trivial ($\nu_- = \nu_+ = 0$) and non-trivial ($\nu_- = -\nu_+ = 1$) phases of the system, respectively. The width of the bandgap is given by $\Delta_g = 2|\delta| = 2|t_1 - t_2|$, and the ranges of two bands are $\Delta_{\varepsilon 1} = \Delta_{\varepsilon 2} = (t_1 + t_2) + \delta = 2\max(t_1, t_2)$. When $\delta < 0$, the SSH lattice supports two topological edge states with close to zero eigenvalues, which can be numerically or asymptotically calculated[46].

Considering the two bands as two subsystems, the optical power and internal energy of each subsystem can be calculated by

$$P_1 = \sum_{i=1}^{M_1} n_i, \quad P_2 = \sum_{i=M-M_2+1}^{M} n_i, \quad U_1 = \sum_{i=1}^{M_1} \varepsilon_i n_i, \quad U_2 = \sum_{i=M-M_2+1}^{M} \varepsilon_i n_i, \tag{8}$$

where $M_1$ and $M_2$ are the numbers of modes of the lower band and upper band, respectively. In a trivial lattice, $M_1 = M_2 = M/2$, and in the nontrivial system $M_1 = M_2 = M/2 - 1$.

By applying the equation of state (5), we can uniquely determine the thermal equilibrium values of the optical temperature $T_{th}$ and the chemical potential $\mu_{th}$ of the lattice via the following relation[47]

$$\sum_i^M \frac{T_{th}}{P\varepsilon_i - U + T_{th}M} = 1. \tag{9}$$

In this paper we will focus on prethermalization phenomena due to the presence of subsystems. Thus, it is important to be able to dynamically and individually characterize each band of the band structure. Numerically, in a similar fashion to Eq. (9), we can continuously track the optical temperature $T_j(Z)$ and the chemical potential $\mu_j(Z)$ of each prethermalized subsystem. This can be done by modifying the summation in Eq. (9) to the modes of each band:

$$\sum_{i=1}^{M_1} \frac{T_1(Z)}{P_1(Z)\varepsilon_i - U_1(Z) + T_1(Z)M_1} = 1, \quad \sum_{i=M-M_2+1}^{M} \frac{T_2(Z)}{P_2(Z)\varepsilon_i - U_2(Z) + T_2(Z)M_2} = 1. \tag{10}$$

From $(T_j(Z), \mu_j(Z))$ we can also compute the associated RJ distribution of each band in isolation:

$$n_i^{RJ1}(Z) = \frac{T_1(Z)}{\varepsilon_i - \mu_1(Z)}, \quad n_i^{RJ2}(Z) = \frac{T_2(Z)}{\varepsilon_i - \mu_2(Z)}. \tag{11}$$

We consider the system to be in a prethermal stage if the above expressions are in a good agreement with the numerical results. If both subsystems reach the same optical temperatures and chemical potentials, then the prethermal RJ distributions and the final thermal RJ distribution are equalized, and the system has reached thermal equilibrium. With the above definition of prethermalization, we see that the prethermal state continuously evolves according to the predicted values of $(T_j(z), \mu_j(z))$. Thus, in contrast to previous works[24, 35, 36], we observe that in multiband systems there is not a single stationary prethermal state but a dynamically evolving one.

To measure and quantify thermalization effects, we utilize the Kullback-Leibler divergence (KLD) (also known as the relative entropy). It is a useful measure in statistical mechanics[48] that quantifies the difference between two distributions, $p_i$ and $q_i$, that is defined as

$$D_{KL}(p_i|q_i) = \sum_{i=1}^{M} n_i \log\left(\frac{p_i}{q_i}\right). \tag{12}$$

In our case, $q_i = n_i^{RJ}$ is the theoretically predicted RJ distribution and $p_i = \langle n_i(Z) \rangle$ is the ensemble average at distance $Z$. The closer the two distributions are to each other, the smaller the KLD value is[49, 50]. We have previously utilized the KLD to quantify the onset of thermalization in different lattices[34]. Note that Eq. (12) can be utilized to measure both thermalization as well as prethermalization. In the latter case, we account for only the power occupation numbers of each band, whereas the theoretical predictions for $n_i^{RJ}$ are determined from the dynamic calculation (as a function of $Z$) of the power and the energy associated with each band. The KLD is positive except if the two distributions are equal, in which case it becomes zero, and additive for independent distributions. A decreasing value of the KLD, indicates that the system is approaching thermalization. Since KLD is an extensive measure, we define the intensive or mean KLD (MKLD) of the entire system $d_{KL}(Z) = D_{KL}(Z)/M$ and the two subsystems $d_{KLj}(Z) = D_{KLj}(Z)/M_j$ to make the comparison between thermalization and prethermalizations processes of different sizes comparable.

### 3. Different stages of the thermalization process

In this work, we focus our attention on investigating the thermalization stages and processes of multiband systems with or without in-band topological edge states. We focus on the particular case of a finite-sized 1D SSH lattice structure both in the topologically trivial and nontrivial regimes. During propagation we keep track to the mean energies and powers of the two subsystems $(U_j(Z), P_j(Z))$. For the total system $(U(Z), P(Z))$ the power is constant and the linear part of the energy exhibits small fluctuations due to nonlinearity. From $(U, P)$ utilizing Eq. (10) we can compute $(T, \mu)$ and thus the power occupation numbers for the final thermal state. On the other hand, from the respective values of the subsystems, we can dynamically (as a function of $Z$) compute temperatures and the chemical potentials of the two subsystems $T_j(Z), \mu_j(Z)$ and the associated RJ distributions $n_i^{RJ1}(Z)$ and $n_i^{RJ2}(Z)$. As schematically illustrated in Fig. 1(a), the thermalization dynamics are divided into three different stages. In the first stage the system starts from the initial excitation condition which is considered to be out-of-equilibrium up to the point where both bands are prethermalized. We consider the system to be prethermalized if the power occupation numbers match the predicted RJ distributions $n_i^{RJ1}(Z)$ and $n_i^{RJ2}(Z)$. During the second stage, the system remains prethermalized but, dynamically, transitions to the final thermalized state where both prethermal temperatures and chemical potentials match those of the whole system. We point out again, that the prethermal state evolves during propagation. Although there is exchange of energy and power between the two subsystems, due to the relatively large size of the bandgap, this process is significantly slower than the restoration of the prethermal equilibrium. The third and final stage is obtained when the system gets thermalized.

Figure 1(b) illustrates the lattice structure, which is formed with $N = 50$ unit cells, one of them is outlined by the dashed box as an example, with the red and blue dots marking the locations of the two sublattices **a** and **b**. The orange bars connecting adjacent sites represent the hopping between waveguides, with solid ones and dashed ones are for intra-cell and inter-cell couplings $t_1$ and $t_2$ respectively. The largest NN coupling coefficient is set to unity $t_{max} = \max(t_1, t_2) = 1$, to approximately maintain the width of each band to $\Delta\varepsilon_1 = \Delta\varepsilon_2 \cong 2$, while the smaller NN coupling is adjustable with the detuning $\delta = t_1 - t_2$. We accommodate zero boundary conditions in our simulations, accounting for a linear array. By changing the value of the detuning $\delta$, we can calculate a series of eigen spectra by solving the eigen function problem of Eq. (3) which are shown in Fig. 1(c). The band structure exhibits a broadening band gap as the values of the coupling detuning deviates from $\delta = 0$. On the other hand, the topological boundary states emerge inside the band gap at around $\varepsilon = 0$, when the system is in the topological non-trivial regime.

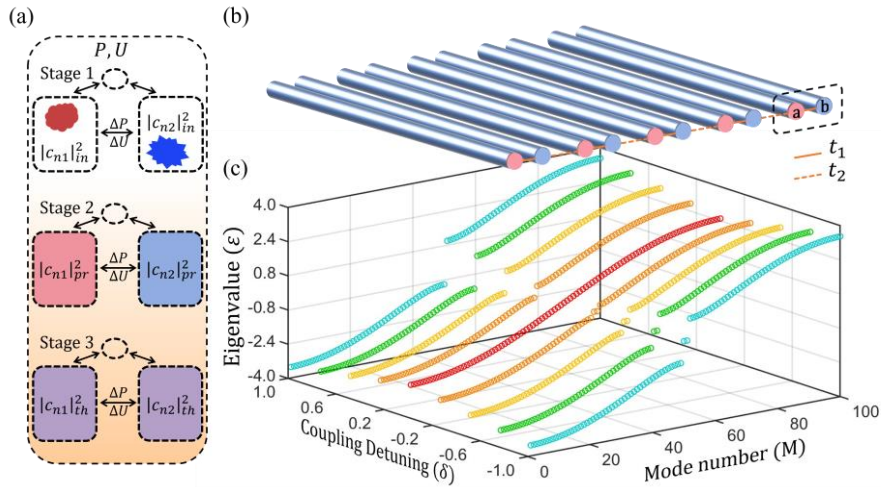

Fig. 1. (a) Schematic illustration with a diagrammatic drawing of the different stages of thermalization. During the thermalization process the two subsystems can almost adiabatically exchange power $P$ and internal energy $U$. In Stage 1, the system evolves from the initial out-of-equilibrium condition to a state where both bands are individually prethermalized. Subsequently, during stage 2 the system transitions to thermalization, meaning that the two prethermalized bands attain a common temperature and chemical potential. This is the third and final stage of the system. (b) Schematic diagram of a finite-sized 1D SSH lattice with intra- and inter-cell coupling coefficients $t_1$ (solid orange lines) and $t_2$ (dash orange lines) between two sub-lattices **a** (red circles) and **b** (blue circles) contained in the unit cell (dash box). (c) Corresponding eigen spectra of a lattice with by $M = 100$ sites and zero boundary conditions vs the detuning between the intra-cell and inter-cell coupling coefficients $\delta = t_1 - t_2$, while the maximum coupling is fixed at $t_{max} = \max(t_1, t_2) = 1$. With $\delta = 0$, the band structure is continuous without band gap (red circles). While $\delta > 0$, the band gaps are opened and broadened with increasing $\delta$ (spectral colored circles behind the red ones). On the other hand, while $\delta < 0$, the system is topological non-trivial with the near zero topological boundary states appearing in the gap (colored circles in front of the red ones).

## 4. The prethermalization

We have studied in detail the thermalization process in SSH lattices. A typical example is presented in Fig. 2 with gap width $\Delta_g = 1$, Kerr coefficient $\gamma = 0.5$ and topological non-trivial near zero boundary states. Numerically, we solve the tight binding system given by Eq. (2) using the Runge-Kutta method. As the initial condition, simultaneously on each band, a cluster of 10 consecutive eigenmodes is uniformly excited with optical power $P_1(0) = P_2(0) = 10$ in total, and internal energy $U_1(0) = -20.4419, U_2(0) = 8.0261$ [Fig. 2(d1)]. We also examined different excitation conditions, and observed qualitatively similar results (see Supplementary Material). The MKLDs are calculated for the average power occupation numbers over 200 random realizations of the entire system $\langle n_i(Z) \rangle$ and for both bands $\langle n_i(Z) \rangle_1$, $\langle n_i(Z) \rangle_2$, where the subscript outside the brackets denotes summation of modes belonging to the 1st or 2nd band. These values are compared with the predicted prethermalization RJ distributions $n_i^{RJ1}(Z)$, $n_i^{RJ2}(Z)$ from the time-varying variables $U_j(Z), U_j(Z)$ or the final R-J distribution $n_i^{RJ}$ by applying Eq. (12). In Fig. 2(a), the MKLDs are illustrated to have decreasing trends as functions of the propagation distance $Z$. We select a threshold value for the MKLD $d_{th} = 5 \times 10^{-4}$ to indicate that the system has thermalized or prethermalized. For example, in Fig. 2(a), we observe that $d_{KL}$ oscillates around this value when the system is thermalized or the subsystems are prethermalized. The values of $d_{KL1}$ and $d_{KL2}$ for the bottom and the upper bands are marked by the red and blue curves respectively, firstly drop and cross the line of the selected common threshold value $d_{th}$ (black dash line) almost at the same distance (around $Z = 1566$) which means both bands get prethermalized simultaneously. Then, as both red and blue curves have small values of the MKLD around the threshold, the quasi-adiabatic subsystems maintain their prethermalization although optical power and internal energy are exchanging between them. After a longer propagation distance, at $Z \cong 7378$, the violet curve for the MKLD of the entire system in the same plot decreases to the threshold value, marking that the entire system settles in the final thermal equilibrium.

More information about this process is obtained by observing the evolution of optical powers (left $y$-axis) and internal energies (right $y$-axis) for both subsystems as well as the entire system. As we can see in Fig. 2(b), the internal energies of the upper and lower bands (blue and red solid curves) will rapidly evolve to their equilibrium values in a very short propagation distance (at $Z \cong 30$). Meanwhile, the optical power distributed on the two bands (blue and red dashed curves) will slowly exchange between each other, showing that the internal energy exchange between two bands is a faster process. The (linear) internal energy of the entire system remains quasi-constant during propagation. Furthermore, in Fig. 2(c), the predicted temperatures of the lower and the upper bands ($T_1$ and $T_2$) are different initially (according to the optical powers and internal energies distributed on two bands), but very rapidly evolve to a common value within the same very short propagation distance $Z \cong 30$. This is expected since temperature is a measure of the energy exchange between subsystems. At a later stage, this common temperature gradually evolves to the predicted temperature of the entire system (violet solid curve). The chemical potentials for the two bands (red and blue dash curves, colors mark the same band as above) are also different at the input. However, their values evolve in a much slower fashion to approach each other. Eventually, they converge when they reach the value of the chemical potential of the entire system, meaning that the system has reached thermal equilibrium (violet dash curve). Since the chemical potential is the conjugate variable of the power, the prethermal values are equalized when there is no mean power exchange between the two bands. Fig. 2(d) provides typical snapshots of the power occupation numbers through the thermalization process. In particular, Fig. 2(d1) shows the initial excitation condition (gray line) at $Z = 0$. Propagating to $Z = 1000$ shown in Fig. 2(d2), the optical occupations have evolved from the input condition but still haven't reached the prethermalization RJ distributions (which also evolve due to the optical power and internal energy exchange). After further propagation, at $Z = 1566$, shown in Fig. 2(d3), the power occupations start to fit the prethermalization RJ distributions. Propagating further, as shown in Fig. 2(d4), the predicted RJ prethermalization distributions have developed to approach the final thermal state. Due to the feeble energy and power exchange between the two subsystems, the SSH lattice maintains its prethermalization in both bands. After long enough propagation, [Fig. 2(d5)], the prethermalization RJ distributions have advanced and reach the final thermalization RJ distribution, with the power occupation numbers also following the final thermal equilibrium state. Again, considering the subsystems as quasi-adiabatic, the power occupation numbers always follow the states of prethermalization, however, the RJ laws of prethermalization themselves evolve towards the final thermalization.

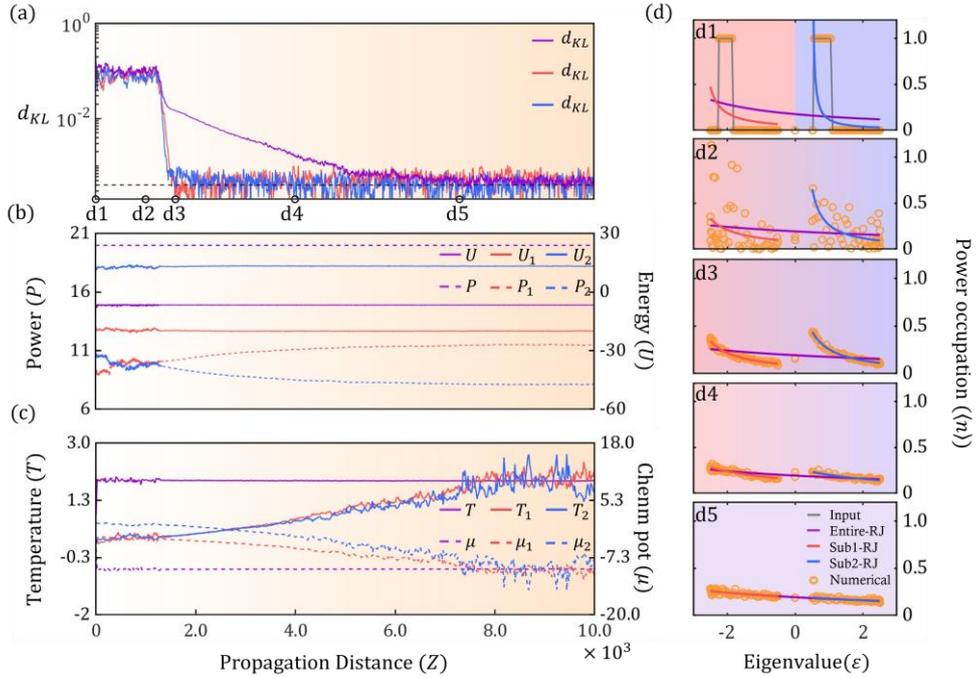

Fig. 2. (a) Mean Kullback-Leibler divergence $d_{KL}$ as a function of propagation distance Z. The $d_{KL}$ for both lower band (blue curve) and upper band (red curve) decrease below the threshold value $d_{th} =$

$5 \times 10^{-4}$ at d3, where both subsystems become prethermalized. On the other hand, the entire system (violet curve) gets final thermalization after d5. (b) The internal energies of the two subsystems $U_1$ and $U_2$ rapidly converge to the thermal equilibrium values (red and blue solid curves), whereas the optical powers for the two the bands $P_1$ and $P_2$ (red and blue dash curves) approach their equilibrium values at a much slower pace. The total power $P$ is conserved while the total energy $U$ exhibits small fluctuations due to nonlinear corrections. (c) The temperatures $T_1$, $T_2$ (solid curves) and the chemical potentials $\mu_1$, $\mu_2$ (dash curves) of the two bands (red and blue curves) evolve towards their thermal equilibrium values of the entire system $T$ and $\mu$ (violet curves). (d) Numerical simulated average power occupations over 200 random realizations (orange circles) at selected $Z$ under uniform initial excitation (grey line) with random phase distribution of the power $P = 10$ over 10 consecutive modes in each band, compared with the theoretical prethermal RJ distributions (red and blue curves) and the final thermal RJ distribution (violet curve). The predicted prethermalization RJ distributions are calculated from the conservation laws $P$ and $U$ of the respective subsystems at the corresponding propagation distances: (d1) $Z = 0$; (d2) $Z = 1000$; (d3) $Z = 1566$; (d4) $Z = 4000$; and (d5) $Z = 7278$. Power occupation curves keep fitting evolutionary prethermalization RJ distributions after d3, then evolve and approach the final RJ distribution of the entire system.

## 5. Effects of topology and bandgap size on thermalization dynamics

Using the MKLD as a measure, we investigate the thermalization dynamics as a function of the band gap width both in the topologically trivial and nontrivial regimes. We have computed the propagation distances where the MKLDs of both subsystems and the entire system firstly reach the threshold value $d_{th} = 5 \times 10^{-4}$, which are regarded as the distances where the system reaches prethermalization or thermalization. First, in Fig. 3(a1, a2), we depict the dynamics of the MKLDs for light propagating in both topological trivial and non-trivial 1D SSH lattices with band gap width $\Delta_g = 1$ and Kerr coefficient $\gamma = 0.5$ (Fig. 3(a2) is the same as Fig. 2(a)). The prethermalization and thermalization distances are marked where the curves of MKLDs firstly cross the threshold line. From that, we can clearly observe the separation between the prethermalization and thermalization distances. We see that in both the trivial and the nontrivial regime and for both bands, the associated four prethermalization distances are almost the same due to the similar excitation conditions. In contrast, the thermalization distances are critically influenced from the presence of the mid-gap boundary states leading to significantly smaller thermalization distances in the topologically non-trivial regime.

For a fair comparison, we keep the width of each band almost constant $\Delta_{b1} = \Delta_{b2} \cong 2$, and adjust the detuning between the intra- and inter-cell coupling coefficients $\delta = t_1 - t_2$ in a lattice with $N = 50$ primitive cells. Note that by changing the sign of the detuning $\delta$ we change the topological properties of the lattice. With the same initial condition as in last section, in Fig. 3(b), we present the prethermalization and thermalization distances with and without the topological boundary states, computed for eight different sizes of the topological band gap in the presence of self-focusing nonlinearity $\gamma = 0.5$. It is apparent from the figure that, the final thermalization of the total system is accelerated by narrowing the width of the band gap. We also find out that when the band gap width narrows down to around $\Delta_g = 0.5$, the prethermalization and thermalization distances merge together. On the other hand, by increasing the width of the bandgap, the thermalization distance of the trivial system increases significantly faster as compared to the nontrivial case This means that the appearance of a prethermalization stage in the thermalization process demands a wide enough band gap. We see that the prethermalization stage exist even in the presence of the mid-gap topological boundary states. Comparing the thermalization distances between the topological trivial and nontrivial SSH lattice, we see that under the same conditions, the systems with topological boundary states thermalizes significantly faster than the trivial ones. Thus, the mid-gap boundary states bridge and accelerate the thermal interaction between the two bands. On the contrary, the prethermalization distances are hardly influenced by either the band gap width or the existence of topological boundary states, which corroborate that due to the quasi-adiabatic band interactions, the thermal exchange between the subsystems is negligible as compared to inner-band interactions. However, the influence on thermalization by a mid-gap non-trivial topological mode vs. an in-gap trivial defect mode, especially in the case of 2D Chern insulators, merits further study.

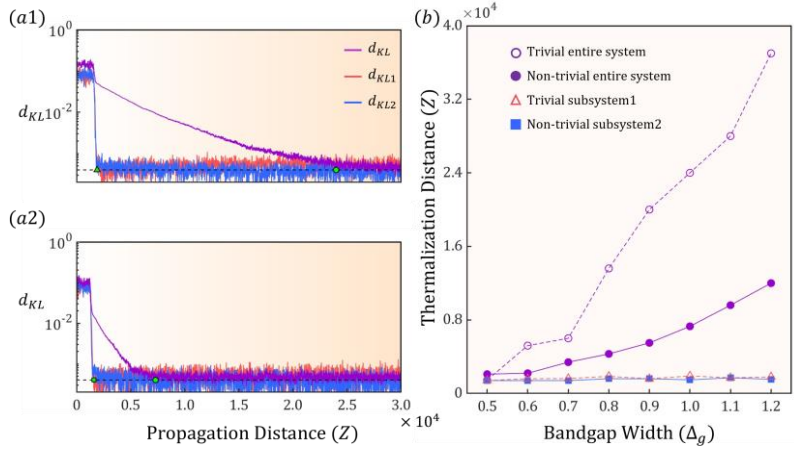

Fig. 3. (a) Dynamics of the mean Kullback-Leibler divergence $d_{KL}$ for (a1) a topological trivial and (a2) a topologically non-trivial 1D SSH lattice. The MKLDs for the subsystems (red and blue curves) approach the threshold value $d_{KL} = 5 \times 10^{-4}$, which is marked with the black dashed line at almost the same propagation distances (green triangle and square), indicating that the system starts to settle into the prethermalization. Then it finally becomes thermalized as the MKLD of the entire system (violet curve) reaches the same threshold (marked by the green circles). (b) Variation of the prethermalization (red and blue curves) and thermalization distances (violet curve) $Z$ for different band gap sizes $\Delta_g$ for one of the subsystems and the entire system in the topological trivial (solid) and non-trivial (dotted) regime with $M = 100$ and maximum coupling coefficient fixed at $t_{max} = 1$.

## 6. Effects of nonlinearity and temperature on thermalization dynamics

In this section, the effects of the strength of the nonlinearity and the optical temperatures of the system and subsystems in the thermalization process are investigated. First, we keep the same initial excitation conditions as in the last section, select a band gap size $\Delta_g \cong 1$, and vary the values of the Kerr nonlinear coefficients $\gamma$. In our numerical simulations, we keep the strength of the nonlinearity to relatively small values to avoid the formation of optical solitons during propagation. As seen in Fig. 4(a), increasing the strength of the nonlinearity $\gamma$ leads to faster processes for both the prethermalization and the thermalization stages. This is an expected result since wave mixing terms becomes stronger with increasing $\gamma$.

In Fig. 4(b) we present the thermalization distances for both trivial and nontrivial systems with Kerr type nonlinearity $\gamma = 0.5$, by simultaneously shifting the excitation positions of the 10 uniformly excited consecutive eigenmodes on both bands, leading to optical temperatures for the entire systems ranging from $T \cong 0.71$ to $T \cong 3.35$. The thermalization process is accelerated as the optical temperature $T$ increases. The synchronous change of the temperatures for the subsystems from $T_{sub1} =$

$T_{sub2} \cong 0.04$ to $T_{sub1} = T_{sub2} \cong 0.45$ will also slightly influence the prethermalization in the same manner.

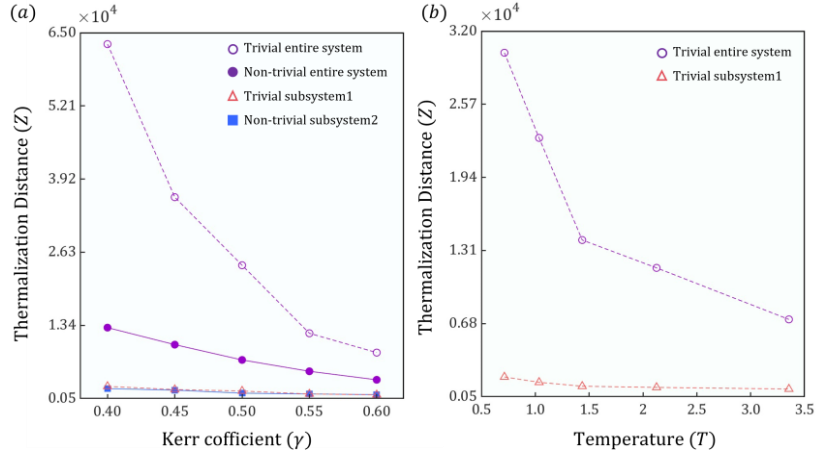

Fig. 4. (a) Variation of the prethermalization and thermalization distances $Z$ as a function of the Kerr nonlinear coefficients $\gamma$, for the two subsystems (red and blue curves) and the entire system (violet curve) in both the topological trivial (solid) and the non-trivial (dotted) regimes. (b) Prethermalization and thermalization distances $Z$ for one of the subsystems (red curve) and the entire system (violet curve) are expedited simultaneously as the tunable excitation conditions provide higher temperatures for both the entire system (from $T \cong 0.7$ to $T \cong 3.35$) and the subsystems (from $T_{sub} \cong 0.04$ to $T_{sub} \cong 0.45$) by shifting the position of the two clusters of 10 uniformly excited consecutive super-modes in modal space.

## 7. Conclusions

In conclusion, we have explored the thermalization process in multiband systems using the one-dimensional Su-Schrieffer-Heeger (SSH) lattice as the simplest topological model. By employing the mean Kullback-Leibler divergence, we have uncovered the distinct stages of thermalization in such systems. Our analysis has also considered factors such as the influence of topological band gap boundary states, gap size, nonlinearity, and optical temperature on the thermalization dynamics. This investigation establishes a theoretical framework for understanding prethermalization and thermalization processes in topological systems. Through the using of this minimal model, we gain insight into how such systems approach thermal equilibrium. Beyond simply advancing our theoretical understanding of thermalization in topological systems, our findings may have practical implications for high power beam delivery via thermal processes that can control the optical temperature and thus the modal distribution across a range of applications in photonics and materials sciences. While the present work is theoretical in nature, our findings are experimentally accessible and can be observed in various photonic lattice platforms, including laser-written waveguide arrays, photorefractive crystals, and multicore fibers. For example, the mode distribution at the output of these systems can be measured and analyzed to verify the predicted prethermalization and thermalization dynamics. Experimental efforts to this end are currently underway. We anticipate that our results will contribute to the emerging field of nonlinear optical thermodynamics, offering valuable perspectives for manipulating thermal dynamics in multiband topological systems and opening new avenues for the design of advanced devices and materials.

**Supplementary Material**

See the supplementary material for simulation results with different initial excitation conditions in the two bands of the SSH lattice.

**Funding**

We acknowledge financial support from the National Key R&D Program of China under Grant (2022YFA1404800), and the National Natural Science Foundation (12134006, 11922408, 12250410236).

**Disclosures**

The authors declare no conflicts of interest.

**Data availability**

Numerical results presented in this manuscript are available from authors upon reasonable request.